\documentclass[12pt]{article}

\catcode`\@=11
\@addtoreset{equation}{section}

\global\arraycolsep=2pt
\usepackage{mathrsfs,amsbsy,amssymb,latexsym,amsfonts,amsmath,cite}

\allowdisplaybreaks

\newcommand\CQ{{\mathcal Q}}

\newcommand\CI{{\mathcal I}}
\newcommand\CJ{{\mathcal J}}

\newcommand\nn{\nonumber}
\newcommand\CN{{\mathcal N}}

\newcommand{\sh}{Schr\"odinger }

\begin{document}

\begin{flushright}
\parbox{4.2cm}
{OIQP-08-05 \hfill \\ 
NSF-KITP-08-82}
\end{flushright}

\vspace*{0.5cm}

\begin{center}
{\Large \bf 
More  super Schr\"odinger algebras from psu(2,2$|$4)
}
\end{center}
\vspace{10mm}

\centerline{\large Makoto Sakaguchi$^{a}$
 and Kentaroh Yoshida$^{b}$
}

\vspace{8mm}

\begin{center}
$^a$ {\it Okayama Institute for Quantum Physics \\
1-9-1 Kyoyama, Okayama 700-0015, Japan} \\
{\tt makoto\_sakaguchi\_at\_pref.okayama.jp}
\vspace{5mm}

$^b$ {\it Kavli Institute for Theoretical Physics, \\ 
University of California, Santa Barbara,  \\
Santa Barbara CA.\ 93106, USA} 
\\
{\tt kyoshida\_at\_kitp.ucsb.edu}
\end{center}

\vspace{1cm}

\begin{abstract} 
We discuss super Schr\"odinger algebras with less supercharges from
$\CN$=4 superconformal algebra psu(2,2$|$4). Firstly $\CN$=2 and $\CN$=1
superconformal algebras are constructed from the psu(2,2$|$4) via
projection operators. Then a super Schr\"odinger subalgebra is found
from each of them. The one obtained from $\CN$=2 has 12 supercharges
with su(2)$^2\times$u(1) and the other from $\CN$=1 has 6
supercharges with u(1)$^3$\,. By construction, those are still
subalgebras of psu(2,2$|$4). Another super Schr\"odinger algebra, which
preserves 6 supercharges with a single u(1) symmetry, is also obtained
from $\CN$=1 superconformal algebra su(2,2$|$1). In particular, it
coincides with the symmetry of $\CN$=2 non-relativistic Chern-Simons
matter system in three dimensions.
\end{abstract}

\thispagestyle{empty}
\setcounter{page}{0}

\newpage

\section{Introduction}

A \sh algebra \cite{Sch1,Sch2} is known as a non-relativistic analog of
relativistic conformal algebra. The algebra with $d$ spatial dimensions
may be embedded into a conformal algebra so($d$+2,2). Hence it is an
interesting attempt to consider a role of the \sh algebra in the context
of AdS/CFT correspondence \cite{AdS/CFT,GKP,W}.

\medskip 

Non-relativistic CFT (NRCFT), which has \sh symmetry, is discussed in
\cite{Henkel,MSW,SW,NS} and it is expected to have an application to a
cold atom system \cite{Son,BM}. A candidate of the gravity dual, which
preserves the \sh symmetry as the maximal one, is proposed in
\cite{Son,BM}. As another scenario, it has been proposed in
\cite{Goldberger,BF} that one may consider AdS/NRCFT without deforming
the metric and including any exotic matters.  It would be an interesting
direction to consider a supersymmetric extension of \cite{Goldberger,BF}
by considering the standard setup of AdS/CFT.

\medskip 

As the first step, super \sh algebras should be found from the
superconformal algebras. We have obtained the \sh algebras with 24
supercharges as subalgebras of psu(2,2$|$4), osp(8$|$4) and
osp(8$^{\ast}|$4) \cite{SY:Sch1}, which preserve 16 rigid
supersymmetries while half of 16 superconformal generators are projected
out. The \sh algebras may be realized in the corresponding gauge
theories. But the field theoretical model, 
which has the super Schr\"odinger symmetry as the maximal one, 
has not been revealed yet.

\medskip

The study of super \sh algebra has a long history and it has been
discussed in some contexts \cite{GGT,BH,DH,HU,LLM}. There are some
models possessing super \sh symmetry as the maximal one, such as a
superparticle \cite{GGT}, super harmonic oscillator \cite{BH} and
non-relativistic super Chern-Simons (CS) matter system in three
dimensions \cite{LLM}\footnote{The non-relativistic CS matter system was
originally constructed by Jackiw and Pi \cite{JP}, and its
supersymmetrization has been done in \cite{LLW}.}.

\medskip

But the number of the conserved supercharges in the models is not so
large and, for example, the non-relativistic CS matter system has
$\CN$=2 in three dimensions. On the other hand, our resulting algebras
contain $\CN$=8 in three dimensions. Hence, in order to find any common
ground between our procedure and the existing result, it would be nice
to look for less supersymmetric \sh subalgebras of psu(2,2$|$4)\,. If we
could find a point of agreement, it might give a clue to discuss the
gauge-theory side.

\medskip 

In this manuscript we discuss this issue and find more super \sh
subalgebras of psu(2,2$|$4). We first construct $\CN$=2 and $\CN$=1
superconformal algebras\footnote{Those are obtained from $\CN$=4 and
hence should be called $\CN$=2$^{\ast}$ and $\CN$=1$^{\ast}$\,,
respectively.  But for simplicity we will omit $\ast$ hereafter.} from
the psu(2,2$|4$) by constructing projection operators. Then a new super
\sh algebra is found from each of them.  The one obtained from $\CN$=2
contains 12 supercharges with su(2)$^2\times$u(1) R-symmetry and the
other from $\CN$=1 has 6 supercharges with u(1)$^3$\,. Another super
Schr\"odinger algebra, which preserves 6 supercharges with a single u(1)
symmetry, is also obtained from $\CN$=1 superconformal algebra
su(2,2$|$1)\,. In particular, it coincides with the symmetry of the $\CN$=2
non-relativistic CS matter system in three dimensions \cite{LLM}.

\medskip 

This manuscript is organized as follows. In section 2, $\CN$=2 and
$\CN$=1 superconformal algebras are derived from the $\CN$=4
superconformal algebra psu(2,2$|$4) via projection operators. In section
3 super \sh algebras are found as subalgebras of $\CN$=2 and $\CN$=1
superconformal algebras.  Section 4 is devoted to a conclusion and
discussions. In the Appendix we briefly summarize our notation and the
relation between psu(2,2$|$4) and $\CN$=4 superconformal generators.

\section{$\CN=2$ and $1$ conformal algebras from psu(2,2$|$4)}

By constructing projection operators, we will obtain $\CN$=2 and $\CN$=1
superconformal algebras from the $\CN$=4 superconformal algebra
described by psu(2,2$|4$) \footnote{ For the relation between
psu(2,2$|$4) and the generators of $\CN$=4 superconformal, see Appendix
A.}.

\subsection{$\CN=4$ superconformal algebra}

We begin with the four-dimensional $\CN$=4 superconformal algebra. The
commutation relations of the bosonic generators are composed of the
AdS$_5$ part and the S$^5$ part. The AdS$_5$ part is given by\footnote{
We suppress trivial (anti-)commutation relations below.
}
\begin{eqnarray}
&&
[\tilde P_\mu,\tilde D]=-\tilde P_\mu\,, \quad 
[\tilde K_\mu,\tilde D]=\tilde K_\mu\,, \quad 
[\tilde P_\mu, \tilde K_\nu]=\frac{1}{2}\tilde J_{\mu\nu}
+\frac{1}{2}\eta_{\mu\nu} \tilde D\,,
\cr&&
[\tilde J_{\mu\nu},\tilde P_\rho]
 =\eta_{\nu\rho}\tilde P_\mu-\eta_{\mu\rho}\tilde P_\nu\,, \quad 
[\tilde J_{\mu\nu},\tilde K_\rho]
 =\eta_{\nu\rho}\tilde K_\mu-\eta_{\mu\rho}\tilde K_\nu\,, \quad 
 \label{conformal}
 \\ &&
[\tilde J_{\mu\nu}, \tilde J_{\rho\sigma}]
=\eta_{\nu\rho}\tilde J_{\mu\sigma}+\mbox{3-terms}\,, \nn 
\end{eqnarray}
and the S$^5$ part is 
\begin{eqnarray}
&&
[P_{a'},P_{b'}]=-J_{a'b'}\,, \quad 
[J_{a'b'},P_{c'}]=\eta_{b'c'}P_{a'}-\eta_{a'c'}P_{b'}\,, \quad 
\cr&&
[J_{a'b'},J_{c'd'}]=
\eta_{b'c'}J_{a'd'}+\mbox{3-terms}\,. \nn 
\end{eqnarray}
Then the (anti-)commutation relations, which contain the fermionic generators
$\tilde Q$ and $\tilde S$\,, are as follows. Those of the bosonic
generators and the fermionic ones are
\begin{eqnarray}
&&
[\tilde P_\mu,\tilde S]=-\frac{1}{2}\tilde Q\Gamma_{\mu 4}~,~~~
[\tilde K_\mu,\tilde Q]=\frac{1}{2}\tilde S\Gamma_{\mu 4}~,~~~
[\tilde D, \tilde Q]=\frac{1}{2}\tilde Q ~,~~~
[\tilde D, \tilde S]=-\frac{1}{2}\tilde S~,
\cr&&
[\tilde J_{\mu\nu}, \tilde Q]=\frac{1}{2}\tilde Q\Gamma_{\mu\nu}~,~~~
[\tilde J_{\mu\nu},\tilde S]=\frac{1}{2}\tilde S\Gamma_{\mu\nu}~,~~~
\nn \\&&
[P_{a'},\tilde Q]=\frac{1}{2}\tilde Q\Gamma_{a'}\CJ i\sigma_2~,~~~
[J_{a'b'},\tilde Q]=\frac{1}{2}\tilde Q\Gamma_{a'b'}~,~~~
\nonumber\\ &&
[P_{a'},\tilde S]=\frac{1}{2}\tilde S\Gamma_{a'}\CJ i\sigma_2~,~~~
[J_{a'b'},\tilde S]=\frac{1}{2}\tilde S\Gamma_{a'b'}~, \nn 
\end{eqnarray}
and those of the fermionic generators are 
\begin{eqnarray}
&& \{\tilde Q^T,\tilde Q\}
=4iC\Gamma^\mu  p_-h_+\tilde P_\mu\,, \quad 
\{\tilde S^T,\tilde S\}
=4iC\Gamma^\mu  p_+h_+\tilde K_\mu\,, \cr&&
\{\tilde Q^T,\tilde S\} 
=iC\Gamma^{\mu\nu}\CI i\sigma_2p_+ h_+\tilde J_{\mu\nu}
+2i C\Gamma^4p_+h_+ \tilde D \nn \\ 
&& \qquad \qquad \qquad +2iC\Gamma^{a'}p_+h_+ P_{a'}
-iC\Gamma^{a'b'}\CJ i\sigma_2 p_+h_+ J_{a'b'}\,. \nn 
\end{eqnarray}
Here $\tilde Q$ are 16 supercharges while $\tilde S$ are 16
superconformal charges.

\medskip 

Next $\CN$=2 and $\CN$=1 superconformal algebras will be obtained from
the psu(2,2$|4$) by constructing projection operators.

\subsection{$\CN=2$ superconformal algebra with su(2)$\mathbf{^2\times}$u(1)}

From now on we shall derive $\CN$=2 superconformal algebra from $\CN$=4
superconformal algebra.

\medskip 

For that purpose, let us introduce a projection operator defined
by\footnote{ Another projection operator $q_+=\frac{1}{2}(1+
\Gamma^{56}i\sigma_2)$ also leads us to the similar result.  In this
case, su(2)$^2\times$u(1) R-symmetry is generated by $\{P_{\bar
a'},J_{\bar a'\bar b'},J_{56}\}$ with $\bar a'=7,8,9$.}
\begin{eqnarray}
q_+=\frac{1}{2}(1+ \Gamma^{5678})\,, \nn 
\end{eqnarray}
and require that 
\begin{eqnarray}
\tilde Q= \tilde Q q_+~,~~~
\tilde S=\tilde S q_+~. \nn
\end{eqnarray}
$\tilde Q$ are 8 supercharges while $\tilde S$ are 8 superconformal charges. 
Then the anti-commutation relations among $\tilde Q$ and $\tilde S$
are reduced to
\begin{eqnarray}
\{\tilde Q^T,\tilde Q\}
&=&4iC\Gamma^\mu  q_+p_-h_+\tilde P_\mu~,
~~~
\{\tilde S^T,\tilde S\}
=4iC\Gamma^\mu  q_+p_+h_+\tilde K_\mu~,
\cr
\{\tilde Q^T,\tilde S\}
&=&iC\Gamma^{\mu\nu}\CI i\sigma_2q_+p_+ h_+\tilde J_{\mu\nu}
+2i C\Gamma^4q_+p_+h_+ \tilde D
\cr&&
+2iC\Gamma^{9}q_+p_+h_+ P_{9}
-iC\Gamma^{\bar a'\bar b'}\CJ i\sigma_2q_+ p_+h_+ J_{\bar a'\bar b'}\,, 
\nn 
\end{eqnarray}
where $\bar a'=5,6,7,8$\,. 

\medskip 

One can find that the following set of the generators, 
\begin{eqnarray}
\{\tilde K_\mu, \tilde S, \tilde D, \tilde J_{\mu\nu},
P_9,J_{\bar a'\bar b'}, \tilde Q, \tilde P_\mu\} 
\end{eqnarray}
forms $\CN=2$ superconformal algebra.
Since $J_{\bar a'\bar b'}$ generates
so(4)$\cong$su(2)$\times $su(2), the R-symmetry is
su(2)$\times$su(2)$\times$u(1) generated by $\{P_9,J_{\bar a'\bar b'}
\}$\,. The commutation relations between the bosonic generators are
\eqref{conformal} and su(2)$\times$su(2)$\times$u(1), while those
between the bosonic generators and $(\tilde Q,\tilde S)$ are
\begin{eqnarray}
&&
[\tilde P_\mu,\tilde S]=-\frac{1}{2}\tilde Q\Gamma_{\mu 4}~,~~~
[\tilde K_\mu,\tilde Q]=\frac{1}{2}\tilde S\Gamma_{\mu 4}~,~~~
[\tilde D, \tilde Q]=\frac{1}{2}\tilde Q ~,~~~
[\tilde D, \tilde S]=-\frac{1}{2}\tilde S\,, 
\cr&&
[\tilde J_{\mu\nu}, \tilde Q]=\frac{1}{2}\tilde Q\Gamma_{\mu\nu}~,~~~
[\tilde J_{\mu\nu},\tilde S]=\frac{1}{2}\tilde S\Gamma_{\mu\nu}~,~~~
\\&&
[P_{9},\tilde Q]=\frac{1}{2}\tilde Q i\sigma_2~,~~~
[J_{\bar a'\bar b'},\tilde Q]=\frac{1}{2}\tilde Q\Gamma_{\bar a'\bar b'}~,~~~
\cr&&
[P_{9},\tilde S]=\frac{1}{2}\tilde Si\sigma_2~,~~~
[J_{\bar a'\bar b'},\tilde S]=\frac{1}{2}\tilde S\Gamma_{\bar a'\bar b'}~.~~~
\label{so(6) fermionic N=2}
\end{eqnarray}

\subsection{$\CN=1$ superconformal algebra  with u(1)$^3$}

Here we derive $\CN$=1 superconformal algebra from the psu(2,2$|$4). 

\medskip 

Let us introduce a 1/4 projection operator defined by
\begin{eqnarray}
q_+=\frac{1}{2}(1+ \Gamma^{56}i\sigma_2)
\frac{1}{2}(1+ \Gamma^{78}i\sigma_2)~,
\end{eqnarray}
and require that 
\begin{eqnarray}
\tilde Q= \tilde Q q_+\,, \qquad 
\tilde S=\tilde S q_+\,. \nn 
\end{eqnarray}
Here $\tilde Q$ are 4 supercharges while $\tilde S$ are 4 superconformal
charges. The anti-commutation relations of $\tilde Q$ and $\tilde S$ are
reduced to
\begin{eqnarray}
\{\tilde Q^T,\tilde Q\}
&=&4iC\Gamma^\mu  q_+p_-h_+\tilde P_\mu\,, \qquad 
\{\tilde S^T,\tilde S\}
=4iC\Gamma^\mu  q_+p_+h_+\tilde K_\mu\,, \nn \\ 
\{\tilde Q^T,\tilde S\}
&=&iC\Gamma^{\mu\nu}\CI i\sigma_2q_+p_+ h_+\tilde J_{\mu\nu}
+2i C\Gamma^4q_+p_+h_+ \tilde D
-2iC \CJ q_+ p_+h_+ R\,, \nn 
\end{eqnarray}
where we have relabeled the three generators as follows: 
\[
R=R_1+R_2+R_3~,~~~
R_I=(P_9,J_{56},J_{78})\,.
\] 

\medskip 

Then one finds that the set of the generators,
\begin{eqnarray}
\{\tilde K_\mu, ~\tilde S, ~\tilde D, ~\tilde J_{\mu\nu},~
R_I,~ \tilde Q,~ \tilde P_\mu\}\,, \nn 
\end{eqnarray}
forms $\CN$=1 superconformal algebra. Here $R_I$ generate the R-symmetry
u(1)$^3$\,. The commutation relations of the bosonic generators are
written in \eqref{conformal}. Those of the bosonic generators and
$(\tilde Q,\tilde S)$ are given by
\begin{eqnarray}
&&
[\tilde P_\mu,\tilde S]=-\frac{1}{2}\tilde Q\Gamma_{\mu 4}\,,~~~
[\tilde K_\mu,\tilde Q]=\frac{1}{2}\tilde S\Gamma_{\mu 4}\,,~~~
[\tilde D, \tilde Q]=\frac{1}{2}\tilde\,,~~~
[\tilde D, \tilde S]=-\frac{1}{2}\tilde S\,, 
\cr&&
[\tilde J_{\mu\nu}, \tilde Q]=\frac{1}{2}\tilde Q\Gamma_{\mu\nu}\,,~~~
[\tilde J_{\mu\nu},\tilde S]=\frac{1}{2}\tilde S\Gamma_{\mu\nu}\,,~~~ 
\\&&
[R_I,\tilde Q]=-\frac{1}{2}\tilde Q i\sigma_2\,,~~~
[R_I,\tilde S]=-\frac{1}{2}\tilde Si\sigma_2\,.~~~
\label{so(6) fermionic N=1}
\end{eqnarray}
Finally we note that 
\begin{eqnarray}
\{\tilde K_\mu,~
\tilde S,~
\tilde D,~
\tilde J_{\mu\nu},~
R,~
\tilde Q,~
\tilde P_\mu\}
\label{su(2,2|1)}
\end{eqnarray}
forms the superalgebra su(2,2$|$1). 

\section{Super Schr\"odinger algebras}

It is turn to find super \sh subalgebras of $\CN$=2 and
$\CN$=1 superconformal algebras constructed in the previous section.

\medskip

First of all, we consider the bosonic part. The discussion is common in
both $\CN$=2 and $\CN$=1 cases. In order to further reduce the commutation
relations, let us decompose the bosonic generators as follows: 
\begin{eqnarray}
&&
P_\pm=\frac{1}{\sqrt{2}}(\tilde P_0\pm \tilde P_3)~,~~
K_\pm=\frac{1}{\sqrt{2}}(\tilde K_0\pm \tilde K_3)~,~~
J_{i\pm}=\frac{1}{\sqrt{2}}(\tilde J_{i0}\pm \tilde J_{i3})~,~~
\cr&&
D=\frac{1}{2}(\tilde D-J_{03})~,~~
D'=\frac{1}{2}(\tilde D+J_{03})~,~~
P_i=\tilde P_i~,~~
K_i=\tilde K_i~,~~
J_{ij}=\tilde J_{ij} 
\label{decomposition Sch} \\ 
&& \qquad \qquad \mu=(0,i,3), \quad i=1,2. \nn  
\end{eqnarray}
Then it is straightforward to rewrite the commutation relations in
\eqref{conformal} as
\begin{eqnarray}
&&
[J_{ij},J_{k\pm}]=\eta_{jk}J_{i\pm}-\eta_{ik}J_{j\pm}~,~~~
[J_{i\pm},J_{j\mp}]=J_{ij}\pm\eta_{ij}(D'-D)~,~~~
\cr&&
[J_{ij},P_k]=\eta_{jk}P_i-\eta_{ik}P_j~,~~~
[J_{ij},K_k]=\eta_{jk}K_i-\eta_{ik}K_j~,~~~\nn \\ 
&&
[P_i,K_j]=\frac{1}{2}J_{ij}+\frac{1}{2}\eta_{ij}(D'+D)~,~~~
[P_i,K_\pm]=\frac{1}{2}J_{i\pm}~,~~~
[P_\pm, K_i]=-\frac{1}{2}J_{i\pm}~,~~~
\cr&&
[D,J_{i\pm}]=\mp\frac{1}{2}J_{i\pm}~,~~~
[D',J_{i\pm}]=\pm\frac{1}{2}J_{i\pm}~,~~~
\cr&&
[P_i,J_{j\pm}]=\eta_{ij} P_\pm~,~~~
[K_i,J_{j\pm}]=\eta_{ij} K_\pm~,~~~
[J_{i\pm},P_\mp]=-P_i~,~~~
[J_{i\pm},K_\mp]=-K_i~,~~~
\cr&&
[P_+,K_-]=-D'~,~~~
[P_-,K_+]=-D~,~~~
\cr&&
[D,P_-]=P_-~,~~~
[D,P_i]=\frac{1}{2}P_i~,~~~
[D,K_+]=-K_+~,~~~
[D,K_i]=-\frac{1}{2}K_i~,~~~
\cr&&
[D',P_+]=P_+~,~~~
[D',P_i]=\frac{1}{2}P_i~,~~~
[D',K_-]=-K_-~,~~~
[D',K_i]=-\frac{1}{2}K_i~.\nn 
\end{eqnarray}
Here the set of the generators
\begin{eqnarray}
\{J_{ij},~J_{i+},~D, ~P_\pm,~P_i,~K_+\}\,
\label{Sch generators}
\end{eqnarray}
is a subalgebra of so(4,2) and it forms the Schr\"odinger algebra
\begin{eqnarray}
&&
[J_{ij},J_{k+}]=\eta_{jk}J_{i+}-\eta_{ik}J_{j+}~,~~~
[J_{ij},P_k]=\eta_{jk}P_i-\eta_{ik}P_j~,~~~
[J_{i+},P_-]=-P_i~,~~~
 \nn \\ 
&&
[P_i,K_+]=\frac{1}{2}J_{i+}~,~~~
[P_i,J_{j+}]=\eta_{ij} P_+~,~~~
[P_-,K_+]=- D~,~~~
\label{Sch} \\ 
&&
[D,J_{i+}]=-\frac{1}{2}J_{i+}~,~~~
[D,P_-]=P_-~,~~~
[D,P_i]=\frac{1}{2}P_i~,~~~
[D,K_+]=-K_+~. \nn 
\end{eqnarray}
We note that $P_+$ is a center of the \sh algebra.

\medskip 

The remaining problem is the fermionic part, and hereafter we will
discuss it by following the procedure developed in \cite{SY:Sch1}.

\subsection{From $\CN=2$ conformal algebra to super Schr\"odinger algebra}

We consider the $\CN$=2 superconformal algebra in this section.

\medskip 

According to the decomposition \eqref{decomposition Sch}, we rewrite the
(anti-)commutation relations including the fermionic generators as
follows:
\begin{eqnarray}
&& \hspace*{-0.1cm} 
\{\tilde Q^T,\tilde Q\}=
4iC\Gamma^+q_+p_-h_+P_+
+4iC\Gamma^-q_+p_-h_+P_-
+4iC\Gamma^iq_+p_-h_+P_i\,,
\cr&& \hspace*{-0.1cm}
\{\tilde S^T,\tilde S\}=
4iC\Gamma^+q_+p_+h_+K_+
+4iC\Gamma^-q_+p_+h_+K_-
+4iC\Gamma^iq_+p_+h_+K_i\,,
\cr&& \hspace*{-0.1cm}
\{\tilde Q^T,\tilde S\}=
iC\Gamma^{ij}\CI i\sigma_2 q_+p_+h_+ J_{ij}
+2iC\Gamma^{i+}\CI i\sigma_2 q_+p_+h_+ J_{i+}
+2iC\Gamma^{i-}\CI i\sigma_2 q_+p_+h_+ J_{i-}
\cr&&~~~~~~~~~~~~
-2iC\Gamma^4\Gamma^{+}\Gamma^-q_+p_+h_+D'
-2iC\Gamma^4\Gamma^{-}\Gamma^+ q_+p_+h_+D
\cr&&~~~~~~~~~~~~
+2iC\Gamma^{9}q_+p_+h_+ P_{9}
-iC\Gamma^{\bar a'\bar b'}\CJ i\sigma_2 q_+p_+h_+ J_{\bar a'\bar b'}\,,
\cr&& \hspace*{-0.1cm}
[K_\pm,\tilde Q]=-\frac{1}{2}\tilde S\Gamma^{\mp}\Gamma_4\,,~~~
[K_i,\tilde Q]=\frac{1}{2}\tilde S \Gamma_{i4}\,,~~~
[P_\pm,\tilde S]=\frac{1}{2}\tilde Q\Gamma^{\mp}\Gamma_4\,,~~~
[P_i,\tilde S]=-\frac{1}{2}\tilde Q \Gamma_{i4}\,,~~~
\cr&& \hspace*{-0.1cm}
[J_{ij},\tilde Q]=\frac{1}{2}\tilde Q \Gamma_{ij}\,,~~~
[J_{ij},\tilde S]=\frac{1}{2}\tilde S \Gamma_{ij}\,,~~~
[J_{i\pm},\tilde Q]=-\frac{1}{2}\tilde Q \Gamma_i\Gamma^\mp\,,~~~
[J_{i\pm},\tilde S]=-\frac{1}{2}\tilde S \Gamma_i \Gamma^\mp\,,
\cr&& \hspace*{-0.1cm}
[D,\tilde Q]=-\frac{1}{4}\tilde  Q \Gamma^{+}\Gamma^-\,,~~
[D,\tilde S]=\frac{1}{4}\tilde  S \Gamma^-\Gamma^+\,,~~
[D',\tilde Q]=-\frac{1}{4}\tilde  Q \Gamma^-\Gamma^+\,,~~
[D',\tilde S]=\frac{1}{4}\tilde  S \Gamma^+\Gamma^-\,, \nn 
\end{eqnarray}
and \eqref{so(6) fermionic N=2}, where we have defined
$\Gamma^\pm=\frac{1}{\sqrt{2}}(\Gamma^0\pm\Gamma^3)$\,.

\medskip

As was seen above, the set of the generators 
\eqref{Sch generators}
forms the Schr\"odinger algebra. Then we derive a
super-Schr\"odinger algebra from $\CN$=2 superconformal algebra below.

\medskip 

Let us introduce the light-cone projection operator 
\begin{eqnarray}
\ell_\pm=\frac{1}{2}(1\pm \Gamma^{03})=
-\frac{1}{2}\Gamma^\pm\Gamma^{\mp}\,,  
\label{light cone projector}
\end{eqnarray}
which commutes with projectors $h_+$, $p_\pm$ and $q_\pm$,
and decompose $\tilde S$ as
\begin{eqnarray}
S=\tilde S \ell_- ~,~~~
S'= \tilde S \ell_+\,.
\label{decomposition S}
\end{eqnarray}
Then we show that
\begin{eqnarray}
\{J_{ij},~J_{i+},~D, ~P_\pm,~P_i,~K_+,
~\tilde Q, ~S,~ P_9,~J_{\bar a'\bar b'}\}
\label{Sch generators from N=2}
\end{eqnarray}
forms a super Schr\"odinger algebra.

\medskip 

First the anti-commutation relations between
$\tilde Q$ and $S$ are given by 
\begin{eqnarray}
&&\{\tilde Q^T,\tilde Q\}=
4iC\Gamma^+
q_+p_-h_+P_+
+4iC\Gamma^-
q_+p_-h_+P_-
+4iC\Gamma^i
q_+p_-h_+P_i~,
\cr&&
\{ S^T, S\}=
4iC\Gamma^+\ell_- q_+p_+h_+K_+
~, \label{QQ from N=2}
\\ &&
\{\tilde Q^T,S\}=
iC\Gamma^{ij}\CI i\sigma_2 \ell_- q_+p_+h_+ J_{ij}
+2iC\Gamma^{i+}\CI i\sigma_2 \ell_- q_+p_+h_+ J_{i+}
\cr&&~~~~~~~~~~~~
-2iC\Gamma^4\Gamma^{-}\Gamma^+ \ell_- q_+p_+h_+D
\cr &&~~~~~~~~~~~~
+2iC\Gamma^{9}\ell_- q_+p_+h_+ P_{9}
-iC\Gamma^{\bar a'\bar b'}\CJ 
i\sigma_2 \ell_- q_+p_+h_+ J_{\bar a'\bar b'}\,,  \nn 
\end{eqnarray}
where we have used $\Gamma^\pm\ell_\pm=0~$.
In the right-hand sides, the bosonic generators contained in \eqref{Sch
generators from N=2} appear.

\medskip 

Next we examine commutation relations between the bosonic generators in
\eqref{Sch generators from N=2} and $(\tilde Q, S)$
\begin{eqnarray}
&&
[K_+,\tilde Q]=-\frac{1}{2}S\Gamma^{-}\Gamma_4~,~~~
[P_-,S]=\frac{1}{2}\tilde Q\ell_+\Gamma^{+}\Gamma_4~,~~~
[P_i, S]=-\frac{1}{2}\tilde Q\ell_- \Gamma_{i4} ~,~~~
\cr&&
[J_{ij},\tilde Q]=\frac{1}{2}\tilde Q \Gamma_{ij} ~,~~~
[J_{ij}, S]=\frac{1}{2} S \Gamma_{ij} ~,~~~
[J_{i+},\tilde Q]=-\frac{1}{2}\tilde Q \Gamma_i\Gamma^-~,~~~
\cr&&
[D,\tilde Q]=-\frac{1}{4}\tilde  Q \Gamma^{+}\Gamma^-~,~~~
[D,  S]=\frac{1}{4}   S \Gamma^-\Gamma^+~, 
\label{BQ from N=2}
\\ &&
[P_{9},\tilde Q]=\frac{1}{2}\tilde Q i\sigma_2~,~~~
[J_{\bar a'\bar b'},\tilde Q]=\frac{1}{2}\tilde Q\Gamma_{\bar a'\bar b'}~,~~~
\cr&&
[P_{9},S]=\frac{1}{2} Si\sigma_2~,~~~
[J_{\bar a'\bar b'}, S]=\frac{1}{2} S\Gamma_{\bar a'\bar b'}~.~~~
~ \nn 
\end{eqnarray}
The right-hand sides of the commutation relations above contain $\tilde
Q$ and $S$ only. Thus we find that \eqref{Sch generators from N=2} forms
a super Schr\"odinger algebra. The bosonic subalgebra is a direct sum of
the Schr\"odinger algebra and su(2)$^2\times$u(1). The number of the
supercharges is 12 since we have projected out 1/4 supercharges of 16
fermionic generators of $\CN$=2 superconformal algebra. 

\medskip 

We note that the set of generators, \eqref{Sch generators},
su(2)$^2\times$u(1) generators and $Q=\tilde Q\ell_-$, forms a super \sh
algebra with 4 supercharges. It is still a superalgebra even if there are no
su(2)$^2\times$u(1) generators. Such a superalgebra is a superextension
of the \sh algebra with 4 supercharges.

\subsection{From $\CN=1$ conformal algebra to super Schr\"odinger algebra}

We consider the $\CN$=1 superconformal algebra here. 

\medskip 

Under the decomposition \eqref{decomposition Sch}, the commutation
relations, which include the fermionic generators, are
\begin{eqnarray}
&&\{\tilde Q^{T},\tilde Q\}=
4iC\Gamma^+q_+p_-h_+P_+
+4iC\Gamma^-q_+p_-h_+P_-
+4iC\Gamma^iq_+p_-h_+P_i~,
\cr&&
\{\tilde S^{T},\tilde S\}=
4iC\Gamma^+q_+p_+h_+K_+
+4iC\Gamma^-q_+p_+h_+K_-
+4iC\Gamma^iq_+p_+h_+K_i~,
\cr&&
\{\tilde Q^{T},\tilde S\}=
iC\Gamma^{ij}\CI i\sigma_2 q_+p_+h_+ J_{ij}
+2iC\Gamma^{i+}\CI i\sigma_2 q_+p_+h_+ J_{i+}
+2iC\Gamma^{i-}\CI i\sigma_2 q_+p_+h_+ J_{i-}
\cr&&~~~~~~~~~~~~
-2iC\Gamma^4\Gamma^{+}\Gamma^-q_+p_+h_+D'
-2iC\Gamma^4\Gamma^{-}\Gamma^+ q_+p_+h_+D
\cr&&~~~~~~~~~~~~
-2iC\CJ q_+p_+h_+ R
~,
\cr&&
[K_\pm,\tilde Q]=-\frac{1}{2}\tilde S\Gamma^{\mp}\Gamma_4~,~~~
[K_i,\tilde Q]=\frac{1}{2}\tilde S \Gamma_{i4} ~,~~~
\cr&&
[P_\pm,\tilde S]=\frac{1}{2}\tilde Q\Gamma^{\mp}\Gamma_4~,~~~
[P_i,\tilde S]=-\frac{1}{2}\tilde Q \Gamma_{i4} ~,~~~
\cr&&
[J_{ij},\tilde Q]=\frac{1}{2}\tilde Q \Gamma_{ij} ~,~~~
[J_{ij},\tilde S]=\frac{1}{2}\tilde S \Gamma_{ij} ~,~~~
\cr&&
[J_{i\pm},\tilde Q]=-\frac{1}{2}\tilde Q \Gamma_i\Gamma^\mp~,~~~
[J_{i\pm},\tilde S]=-\frac{1}{2}\tilde S \Gamma_i \Gamma^\mp~,
\cr&&
[D,\tilde Q]=-\frac{1}{4}\tilde  Q \Gamma^{+}\Gamma^-~,~~~
[D,\tilde S]=\frac{1}{4}\tilde  S \Gamma^-\Gamma^+~,~~~
\cr&&
[D',\tilde Q]=-\frac{1}{4}\tilde  Q \Gamma^-\Gamma^+~,~~~
[D',\tilde S]=\frac{1}{4}\tilde  S \Gamma^+\Gamma^-~,
\end{eqnarray}
and \eqref{so(6) fermionic N=1}, where we have defined
$\Gamma^\pm=\frac{1}{\sqrt{2}}(\Gamma^0\pm\Gamma^3)$\,.

\medskip

As was seen above, \eqref{Sch generators} is the Schr\"odinger
algebra. Then we derive a supersymmetric extension of the algebra from
$\CN=1$ superconformal algebra below.

\medskip 

Let us introduce the light-cone projection operator
\eqref{light cone projector}
and decompose  $\tilde S$ as \eqref{decomposition S}.
Then we find that
\begin{eqnarray}
\{J_{ij},~J_{i+},~D, ~P_\pm,~P_i,~K_+,
~\tilde Q, ~S,~
R_I\}
\label{Sch generators from N=1}
\end{eqnarray}
forms a super Schr\"odinger algebra. 

\medskip 

First derive anti-commutation relations between $\tilde Q$ and $S$
\begin{eqnarray}
&&\{\tilde Q^{T},\tilde Q\}=
4iC\Gamma^+
q_+p_-h_+P_+
+4iC\Gamma^-
q_+p_-h_+P_-
+4iC\Gamma^i
q_+p_-h_+P_i~,
\cr&&
\{S^{T}, S\}=
4iC\Gamma^+\ell_- q_+p_+h_+K_+
~,
\cr&&
\{\tilde Q^{T},S\}=
iC\Gamma^{ij}\CI i\sigma_2 \ell_- q_+p_+h_+ J_{ij}
+2iC\Gamma^{i+}\CI i\sigma_2 \ell_- q_+p_+h_+ J_{i+}
\cr&&~~~~~~~~~~~~
-2iC\Gamma^4\Gamma^{-}\Gamma^+ \ell_- q_+p_+h_+D
-2iC\CJ \ell_- q_+p_+h_+ R  \,, \nn 
\end{eqnarray}
where we have used $\Gamma^\pm\ell_\pm=0~$\,. In the right-hand sides,
the bosonic generators in \eqref{Sch generators from N=1} appear.

\medskip 

Next we examine the commutation relations between the bosonic generators
in \eqref{Sch generators from N=1} and $(\tilde Q, S)$
\begin{eqnarray}
&&
[K_+,\tilde Q]=-\frac{1}{2}S\Gamma^{-}\Gamma_4~,~~~
[P_-,S]=\frac{1}{2}\tilde Q\ell_+\Gamma^{+}\Gamma_4~,~~~
[P_i, S]=-\frac{1}{2}\tilde Q\ell_- \Gamma_{i4} ~,~~~
\cr&&
[J_{ij},\tilde Q]=\frac{1}{2}\tilde Q \Gamma_{ij} ~,~~~
[J_{ij}, S]=\frac{1}{2} S \Gamma_{ij} ~,~~~
[J_{i+},\tilde Q]=-\frac{1}{2}\tilde Q \Gamma_i\Gamma^-~,~~~
\cr&&
[D,\tilde Q]=-\frac{1}{4}\tilde  Q \Gamma^{+}\Gamma^-~,~~~
[D,  S]=\frac{1}{4}   S \Gamma^-\Gamma^+~\,, 
\label{comm}
\\ &&
[R_I,\tilde Q]=-\frac{1}{2}\tilde Q i\sigma_2\,, \quad 
[R_I,S]=-\frac{1}{2} S i\sigma_2\,.  \nn
\end{eqnarray}
The right-hand sides of (\ref{comm}) contain $\tilde Q$ and $S$
only. Thus the set of the generators \eqref{Sch generators from N=1}
forms a super Schr\"odinger subalgebra. The bosonic subalgebra is a
direct sum of the Schr\"odinger algebra and u(1)$^3$\,. The number of
the supercharges is 6 since we have projected out 1/4 supercharges of 8
fermionic generators of $\CN$=1 superconformal algebra.

\medskip

We note that the set of generators, \eqref{Sch generators}, $R_I$ and
$Q=\tilde Q\ell_-$, forms a super \sh algebra with 2 supercharges. It is
still a superalgebra even if there are no $R_I$. Such a superalgebra is a
superextension of the \sh algebra with 2 supercharges.

\medskip

If we start from $\CN$=1 superconformal algebra su(2,2$|$1) in
\eqref{su(2,2|1)}, we obtain a super \sh algebra with u(1) R-symmetry
\begin{eqnarray}
\{J_{ij},~
J_{i+},~
D, ~
P_\pm,~
P_i,~
K_+,
~\tilde Q, ~
S,~
R\}~.
\label{Sch generators from su(2,2|1)}
\end{eqnarray}

\subsection*{A relation to non-relativistic CS matter system}

It is worth noting the relation between the algebra \eqref{Sch
generators from su(2,2|1)} and $\CN=2$ super \sh algebra constructed in
\cite{LLM}.

\medskip 

The bosonic subalgebra of \eqref{Sch generators from su(2,2|1)}
coincides with the bosonic part of the superalgebra in \cite{LLM} under
the following identification of the generators, 
\begin{eqnarray}
(J_{ij}, P_-,P_i,J_{i+},P_+,D,K_+,R)=
(J,H,P_i,G_i,N_B+N_F,D,K,N_B-\frac{1}{2}N_F)\,, 
\label{identification bosonic}
\end{eqnarray}
up to trivial scalings of generators. Here the right-hand side
represents the generators used in \cite{LLM}.

\medskip 

Then the next task is to consider the fermionic part of the algebra. Our
supercharges are Majorana-Weyl spinors in $(9+1)$-dimensions satisfying
the Majorana condition $\CQ^c=\CQ$\,, as explained in Appendix.

\medskip 

Note that we may choose the charge conjugation matrix as $C=\Gamma^0$
and then $B=1$\,. It implies that $\Gamma_A$'s are real:
$\Gamma_A^*=\Gamma_A$\,. With this choice, the Majorana condition simply
implies that
$\CQ^*=\CQ$\,. 
Since the projectors $p_\pm$\,, $q_+$ and $\ell_\pm$ are real,
the two-component spinors,
$Q,Q'$ and $S$
are real:
$Q^*=Q$\,, $Q'^*=Q'$ and $S^*=S$ where 
$Q=\tilde Q \ell_-$ and
$Q'=\tilde Q \ell_+$\,.

\medskip 

The supercharges used in \cite{LLM} are complex and hence it is
necessary to convert our two-component real supercharges into
one-component complex supercharges. For this purpose, let us introduce a
pair of projectors
\begin{eqnarray}
k_\pm
=\frac{1}{2}(1\pm i\Gamma^{12})~ \nn 
\end{eqnarray}
and decompose the two-component real spinors as
\begin{eqnarray}
q_1=Q k_+~,~~
q_1'=Qk_-~,~~
q_2=Q' k_-~,~~
q_2'=Q'k_+~,~~
q_3=S k_-~,~~
q_3'=S k_+~.
\nn 
\end{eqnarray}
Noting that $k_\pm$ are complex: $k_\pm^*=k_\mp$\,, the
Majorana condition implies that $q_1$, $q_2$ and $q_3$ are complex
one-component spinors
\begin{eqnarray}
q_1^*=q_1'\,, \qquad q_2^*=q_2'\,, \qquad q_3^*=q_3'\,. \nn 
\end{eqnarray}
With the complex supercharges, the anti-commutation relations are
rewritten as
\begin{eqnarray}
\{q_1,q_1^*\}&=&
4iC\Gamma^+k_-\ell_-q_+p_- h_+ P_+~,
\cr
\{q_2,q_2^*\}&=&
4iC\Gamma^-k_+\ell_+q_+p_- h_+ P_-~,
\cr
\{q_1,q_2^*\}&=&4iC\Gamma^1k_+\ell_+q_+p_-h_+ (P_1-iP_2)~,~~~
\cr
\{q_3,q_3^*\}&=&4iC\Gamma^+ k_-\ell_-q_+p_+h_+ K_+~,
\cr
\{q_1^*, q_3\}&=&2iC\Gamma^1\CI i\sigma_2 k_-\ell_-q_+p_+h_+(J_{1+}+iJ_{2+})~,
\cr
\{q_2^*,q_3\}&=&
iC\Gamma^{ij}\CI i\sigma_2 k_-\ell_-q_+p_+h_+J_{ij}
-2iC\Gamma^4 \Gamma^-\Gamma^+ k_-\ell_-q_+p_+h_+ D
\cr&&
-2iC\Gamma^4\Gamma^-\Gamma^+k_-\ell_-q_+p_+h_+ R~, \nn 
\end{eqnarray}
where we have used $k_\pm=\pm i \Gamma^{12}k_\pm$\,. On the other hand,
the commutation relations including the supercharges are
\begin{eqnarray}
&&
[J_{12},q_1]=-\frac{i}{2}q_1~,~~
[J_{12},q_2]=\frac{i}{2}q_2~,~~
[J_{12},q_3]=\frac{i}{2}q_3~,~~
\cr&&
[D,q_2]=-\frac{1}{4}q_2 \Gamma^+\Gamma^-~,~~
[D,q_3]=\frac{1}{4}q_3 \Gamma^-\Gamma^+~,~~
[J_{1+}-iJ_{2+},q_2]=-q_1 \Gamma_1\Gamma^-~,~~
\cr&&
[P_-,q_3]=\frac{1}{2}q_2\Gamma^+\Gamma_4~,~~
[P_1-iP_2,q_3]=-q_1\Gamma_{14}~,~~
[K_+,q_2]=-\frac{1}{2}q_3\Gamma^-\Gamma_4~,~~
\cr&&
[R,q_1]=-\frac{1}{2}q_1 i\sigma_2~,~~
[R,q_2]=-\frac{1}{2}q_2 i\sigma_2~,~~
[R,q_3]=-\frac{1}{2}q_3 i\sigma_2~. \nn 
\end{eqnarray}
Under the identification \eqref{identification bosonic}, the last three
commutation relations are further rewritten, by noting that $P_+$ is
center of the super \sh algebra, as
\begin{eqnarray}
[N_F,q_I]=\frac{1}{3}q_I i\sigma_2\,, \qquad 
[N_B,q_I]=-\frac{1}{3}q_I i\sigma_2 \qquad (I=1,2,3)\,. \nn 
\end{eqnarray}
Thus we have shown that the above (anti-)commutation relations 
coincide with those of \cite{LLM} under the identification 
\eqref{identification bosonic} and $(q_1,q_2,q_3)=(Q_1,Q_2,F)$\,, 
up to trivial rescalings of generators.

\section{Conclusion and Discussion}

We have found more super \sh subalgebras of psu(2,2$|$4).  First $\CN$=2
and $\CN$=1 superconformal algebras have been constructed from the
psu(2,2$|$4) by constructing projection operators. Then a less
supersymmetric \sh algebra has been found from each of them. The
resulting two superalgebras are as follows: the one preserves 12
supercharges with su(2)$^2\times$u(1) symmetry, and the other preserves
6 supercharges with u(1)$^3$ symmetry.

\medskip

We have also found another super \sh algebra preserving 6 supercharges
with a single u(1) symmetry from su(2,2$|$1)\,. This algebra coincides
with the symmetry of the non-relativistic CS matter system in three
dimensions \cite{LLM}.

\medskip 

It is interesting to look for non-relativistic systems which preserve
super \sh symmetry with 24 and 12 supercharges as the maximal ones
(i.e., new super NRCFTs). Perhaps there would be two possible scenarios.
The one is to reduce a four-dimensional superconformal field theory with
a light-like compactification to a three-dimensional theory, according
to the embedding of the \sh algebra into the superconformal one. 
The other is to take the standard non-relativistic limit 
of certain relativistic models as in \cite{LLM}. 
It would also be
interesting to consider the gravity dual of the non-relativistic CS
matter system.

\medskip 

We hope that our results would be a key to open a new arena to study the
AdS/NRCFT correspondence.

\section*{Acknowledgment}

The authors would like to thank H.~Y.~Chen, S.~Hartnoll, T.~Okuda,
S.~Ryu and A.~Schnyder for useful discussion. The work of MS was
supported in part by the Grant-in-Aid for Scientific Research
No.19540324 from the Ministry of Education, Science and Culture,
Japan. The work of KY was supported in part by JSPS Postdoctoral
Fellowships for Research Abroad and the National Science Foundation
under Grant No.\,NSF PHY05-51164.

\appendix

\section*{Appendix}

\section{psu(2,2$|$4) as $\CN=4$ superconformal algebra}

We briefly explain the relation between psu(2,2$|$4) and the
generators of the $\CN$=4 superconformal algebra. The commutation
relations of the psu(2,2$|$4) are as follows. 
The bosonic part is composed of the so(2,4)
algebra
\begin{eqnarray}
&& [P_a,P_b]=J_{ab}~,~~~
[J_{ab},P_{c}]=\eta_{bc}P_{a}-\eta_{ac}P_{b}~, \nn \\
&& [J_{ab},J_{cd}]=\eta_{bc}J_{ad}+\mbox{3-terms} \qquad
 (a=0,1,2,3,4)\,,  
\end{eqnarray}
and the so(6) algebra
\begin{eqnarray}
&& [P_{a'},P_{b'}]=-J_{a'b'}~,~~~
[J_{a'b'},P_{c'}]=\eta_{b'c'}P_{a'}-\eta_{a'c'}P_{b'}~, \nn \\ 
&& [J_{a'b'},J_{c'd'}]=
\eta_{b'c'}J_{a'd'}+\mbox{3-terms} \qquad (a'=5,6,7,8,9)\,. 
\label{s(6)}
\end{eqnarray} 
The (anti-)commutation relations, which include the fermionic generator $Q$\,, are 
\begin{eqnarray}
&&
[P_a,\CQ]=-\frac{1}{2}\CQ\Gamma_a\CI i\sigma_2~,~~~
[P_{a'},\CQ]=\frac{1}{2}\CQ\Gamma_{a'}i\sigma_2~,~~~
[J_{AB},\CQ]=\frac{1}{2}\CQ\Gamma_{AB}~,\cr
&&
\{\CQ^T,\CQ\}=2iC\Gamma^Ah_+P_A
+iC \Gamma^{ab}\CI i\sigma_2 h_+ J_{ab}
-iC \Gamma^{a'b'}\CJ i\sigma_2 h_+ J_{a'b'}\,, 
 \qquad A=(a,a') \nn \\ 
&& 
\CI=\Gamma^{01234}\,, \quad \CJ=\Gamma^{56789}\,. \nn 
\end{eqnarray}
Here $\Gamma^A$'s are $(9+1)$-dimensional gamma-matrices and $C$ is the
charge conjugation matrix satisfying 
\begin{eqnarray}
\Gamma_A^T=-C\Gamma_AC^{-1}~,~~~
C^\dag C=1~,~~~
C^T=-C~.
\end{eqnarray} 
The supercharge $\CQ$ is a pair of Majorana-Weyl spinors in
$(9+1)$-dimensions. 
The charge conjugation of $\CQ$ is defined by
\begin{eqnarray}
\CQ^c=\CQ^*B^{-1}\,,
\end{eqnarray}
where the matrix $B$ relates $\Gamma_A^*$ and $\Gamma_A$ by
\begin{eqnarray}
\Gamma_A^*=B\Gamma_AB^{-1}~,~~~
B^\dag B=1~,~~~
B^T=B\,.
\end{eqnarray}
It is also related to $C$ via 
\[
C=B\Gamma_0^\dag\,.  
\]
The Majorana-Weyl spinor
$\CQ$ satisfies the Majorana condition 
\[
\CQ^c=\CQ
\] 
as well as the Weyl condition 
\[
\CQ=\CQ h_+\,, \qquad h_+ \equiv \frac{1}{2}(1+\Gamma_{01\cdots 9})~:~
\mbox{chirality projector}\,.
\] 

\medskip

By recombining the generators, we define the followings
\begin{eqnarray}
&&
\tilde P_\mu\equiv \frac{1}{2}(P_\mu-J_{\mu4})~, \quad 
\tilde K_\mu \equiv \frac{1}{2}(P_\mu+J_{\mu4})~, \quad \tilde D \equiv P_4~, \quad 
\tilde J_{\mu\nu}\equiv J_{\mu\nu}~,~~\nn \\
&& \tilde Q\equiv \CQ p_-~, \quad 
\tilde S \equiv\CQ p_+  \qquad\quad (a=(\mu,4),~~\mu=0,1,2,3)\,. 
\nn 
\end{eqnarray}
Here the  projectors $p_\pm$ are defined by
\begin{eqnarray}
p_\pm\equiv\frac{1}{2}(1\pm \Gamma^4\CI i\sigma_2)
=\frac{1}{2}(1\pm \Gamma^{0123}i\sigma_2)~,
\label{p_pm}
\end{eqnarray}
which commute with the chirality projector $h_+$\,.
By noting that 
\[
p^T_\pm C=Cp_\pm\,, \qquad 
\Gamma^{0123}i\sigma_2 p_\pm=\pm p_\pm\,,
\]
the (anti-)commutation relations can also be rewritten as
\begin{eqnarray}
&&
[\tilde P_\mu,\tilde D]=-\tilde P_\mu~,~~~
[\tilde K_\mu,\tilde D]=\tilde K_\mu~,~~~
[\tilde P_\mu, \tilde K_\nu]=\frac{1}{2}\tilde J_{\mu\nu}
+\frac{1}{2}\eta_{\mu\nu} \tilde D~,~~~
\cr&&
[\tilde J_{\mu\nu},\tilde P_\rho]
 =\eta_{\nu\rho}\tilde P_\mu-\eta_{\mu\rho}\tilde P_\nu~,~~~
[\tilde J_{\mu\nu},\tilde K_\rho]
 =\eta_{\nu\rho}\tilde K_\mu-\eta_{\mu\rho}\tilde K_\nu~,~~~\nn \\ 
&& 
[\tilde J_{\mu\nu}, \tilde J_{\rho\sigma}]
=\eta_{\nu\rho}\tilde J_{\mu\sigma}+\mbox{3-terms}~,
\cr&&
\{\tilde Q^T,\tilde Q\}
=4iC\Gamma^\mu  p_-h_+\tilde P_\mu~,
~~~
\{\tilde S^T,\tilde S\}
=4iC\Gamma^\mu  p_+h_+\tilde K_\mu~,
\cr&&
\{\tilde Q^T,\tilde S\}
=iC\Gamma^{\mu\nu}\CI i\sigma_2p_+ h_+\tilde J_{\mu\nu}
+2i C\Gamma^4p_+h_+ \tilde D \nn \\ 
&& \qquad \qquad \qquad +2iC\Gamma^{a'}p_+h_+ P_{a'}
-iC\Gamma^{a'b'}\CJ i\sigma_2 p_+h_+ J_{a'b'}~,
\cr&&
[\tilde P_\mu,\tilde S]=-\frac{1}{2}\tilde Q\Gamma_{\mu 4}~,~~~
[\tilde K_\mu,\tilde Q]=\frac{1}{2}\tilde S\Gamma_{\mu 4}~,~~~
[\tilde D, \tilde Q]=\frac{1}{2}\tilde Q ~,~~~
[\tilde D, \tilde S]=-\frac{1}{2}\tilde S 
\cr&&
[\tilde J_{\mu\nu}, \tilde Q]=\frac{1}{2}\tilde Q\Gamma_{\mu\nu}~,~~~
[\tilde J_{\mu\nu},\tilde S]=\frac{1}{2}\tilde S\Gamma_{\mu\nu}~,~~~\nn
\\&&
[P_{a'},\tilde Q]=\frac{1}{2}\tilde Q\Gamma_{a'}\CJ i\sigma_2~,~~~
[J_{a'b'},\tilde Q]=\frac{1}{2}\tilde Q\Gamma_{a'b'}~,~~~
\cr&&
[P_{a'},\tilde S]=\frac{1}{2}\tilde S\Gamma_{a'}\CJ i\sigma_2~,~~~
[J_{a'b'},\tilde S]=\frac{1}{2}\tilde S\Gamma_{a'b'}\,.~~~\nn 
\end{eqnarray}
The so(6) part is given in \eqref{s(6)}. Thus the resulting algebra is
nothing but the four-dimensional $\CN$=4 superconformal algebra. Here
$\tilde Q$ are 16 supercharges while $\tilde S$ are 16 superconformal
charges.

\end{document}